\documentclass[12pt]{article}
\usepackage{amsfonts,amssymb,amsmath}
\usepackage[dvips]{epsfig}
\begin{document}
\title{\bf Selective transfer of superposition of coherent states by exploiting a cavity QED system}
\author{ N. Behzadi $^{a}$
\thanks{E-mail:n.behzadi@tabrizu.ac.ir}  ,
S. Kazemi Rudsary $^{b}$
\thanks{E-mail:s.kazemirudsary89@ms.tabrizu.ac.ir}
\\ $^a${\small Research Institute for Fundamental Sciences, University of
Tabriz,}
\\ $^b${\small Department of Theoretical Physics
and Astrophysics, Faculty of Physics,}
\\ {\small University of
Tabriz, Tabriz 51666-16471, Iran.}} \maketitle
\begin{abstract}
\noindent We propose a scheme on the basis of a N+2 identical single-mode coupled-cavity QED system for selective transfer of a qubit constructed from superposition of standard coherent states. The cavities arranged in such way that the intermediate or channel cavity is connected uniformly to the sender and N receiver cavities. We consider N different ternary sets of identical QDs whose QDs have been distributed in the sender, channel and one of the receiver cavities respectively. We demonstrate a situation in which the dynamics of the system is confined selectively in a sub sector belongs to one of the ternary set of QDs. This selective dynamics is able to transfer the coherent state-constructed qubit (CSCQ) from the sender party to the desired receiver one reliably. Also, we illustrate that the scheme is optimally robust due to dissipations arises from photon losses in the cavities.
\end{abstract}

\section{Introduction}
One of important requirement of distributed quantum information processing (QIP) in a quantum network is the
ability to establish correlations between arbitrary distant parties in order to perform quantum state transfer (QST) between them. Implementing such protocols in the realm of quantum mechanics, needs to many body interacting quantum
systems like spin chains \cite{bose1} or cavity quantum electrodynamics (QED) systems which each
of them contains an emitter like atom \cite{cirac}.
Coupled cavities not only can be considered as a tool for
observing of quantum cooperative phenomena in strongly
correlated many-body systems \cite{har} but also has potential
applications in QIP \cite{ogd}. Indeed, the system of high-Q
cavities and atoms, or specifically semiconductor quantum dots (QDs) as artificial atoms,
in the strong interaction regime is one of the experimentally realizable systems in which the
intrinsic quantum mechanical coupling dominates losses arisen due to dissipation \cite{cirac, henn, khit}.

Since, in the selective quantum state transfer or quantum routing protocols throughout a quantum network, the initial state is usually prepared at a node, its transmission to another node will inevitably be affected by the detrimental
dispersion of the wave packet hence, high quality state reconstruction will be a hard task. Despite to the quantum state transfer along linear chains, few paper have been devoted to study the selective transfer throughout various quantum networks. In Ref. \cite{david}, routing of quantum information in qubit chains was achieved by suitably chosen
time-dependent local fields acting on the qubits. Multiparty
quantum communication using a spin ring with twisted
boundary conditions provided by a magnetic
flux through the ring was presented in \cite{bose}. The necessity of engineering
of couplings, which becomes more complicated in perfect routing of a quantum state on
an arbitrary path in a regular network with arbitrary spatial dimension, can be removed
by taking quasi-uniform couplings which leads to obtain perfect state transfer in small
regions of a spin network and tailing these regions to obtain perfect state transfer in the
whole system \cite{ross, karim}. It should be noted that it is necessary to study efficient routing protocols that require minimal engineering and external manipulation. At this stage, Paganelli $et$ $al.$ presented two different possible implementations of a router that allows quantum state transfer
from a sender to a chosen receiver by means of a resonant
coupling mechanism \cite{lorenzo}. Also, Ajoy and the coauthor investigated strategies to achieve perfect quantum information transport in arbitrary spin networks \cite{ajoy}.

On the other hand, in all of these protocol based on spin systems, single spin addressing is a hard task because the spatial separation between the neighboring spins is very small \cite{li}. Thus, from the implementation point of view, control over interaction parameters between the spins or over individual spins is almost unfeasibe. However, in a cavity QED system, individual cavities addressing with optical laser is easily achievable. Also, the interaction of a cavity and a semiconductor QD, as an artificial atom, can be engineered in such way that the QD in the cavity has relatively long-lived energy levels  suitable for encoding quantum information \cite{yabu}. By keeping these in mind, quantum computation on the basis of superposition of coherent states as logical qubits (CSCQ), has several advantages \cite{gla, jeo, ral}. Therefore, selective transfer of  quantum information encoded on a typical CSCQ can be regarded as an unavoidable task.

In this paper, we introduce a protocol for selective transfer of quantum information in the form CSCQ using a cavity QED system. We consider N distinct set of QDs for each set, called as a ternary set, there exist three identical QDs distributed in the sender, channel and one of the receiver cavities in such way that each of the receiver cavities contains only one QD. So in this way, there exist N QDs in the sender cavity and N QDs in the channel one. As illustrated in the text, in the sender cavity (unlike to the condition in the channel cavity),  it is assumed that the field mode has interaction only with one of the QDs however, this interaction can be switched on to the other QD. In the presence of suitable detuning between the excitonic modes of various ternary sets and the field mode of the cavities the dynamics of the system is effectively confined in one the ternary sets of QDs. Switching the interaction of the field mode in the sender cavity on to the other QD leads the dynamics of the system takes place in the other ternary set. This is equivalent to the fact that when a general CSCQ is prepared on the QDs selected to interact with the filed mode at the sender party, it can be found reliably on the corresponding QD at the desired receiver party. Finally, since the effective dynamics of the system is lied in one of the ternary sets of QDs then the field mode of all of cavities along with the excitonic modes of other QDs are approximately in their respective vacuum states, suppressing greatly the efficient dissipations arisen from photon and exciton losses during the state transfer processes \cite{iris, zhen, maje}.

This paper is laid out in the following way. In section 2, we describe the basic
properties of the scheme. In Section 3, we are going to realize the selective transfer of a typical CSCQ from the sender party to the desired receiver or target party along with an illustration about the minimal dissipation effects influencing the effective dynamics of the system. The paper is ended with a brief conclusion.

\section{Hamiltonian and dynamics}
We consider a cavity QED system with N+2 identical single-mode cavities in which the central or channel cavity is coupled uniformly to the sender and N receiver cavities via photon hopping process (for simplicity and without lose of generality, we have depicted a cavity system with N=2 receivers in Fig. 1). There exist N different ternary sets of QDs whose QDs have been trapped in sender, channel and one of the receiver cavities in such way that there exist only one QD in each of the receiver cavities (Fig. 1). It is assumed that for each of QDs
there are a few electrons excited from
valance-band to conduction-band and the excitation density
of the Coulomb correlated electron-hole pairs, excitons, in the
ground state is low. Consequently, exciton
operators can be approximated with boson operators. Also, all
nonlinear terms including exciton-exciton interactions and the
phase-space filling effect can be neglected. It is also assumed that
the ground energy of the excitons in each QD is the same.
The Hamiltonian Under the rotating wave approximation is given by
\begin{eqnarray}
&&\hspace{-133mm}\hat{H}_{i}=\hbar \omega_{s,f} \hat{a}_{s}^{\dagger}\hat{a}_{s}+\hbar \omega_{c,f} \hat{a}_{c}^{\dagger}\hat{a}_{c}+\hbar \sum_{j=1}^{N}\omega_{r_{j},f_{j}}\hat{a}_{r_{j}}^{\dagger}\hat{a}_{r_{j}}+\hbar \omega_{s,e_{i}} \hat{b}_{s_{i}}^{\dagger}\hat{b}_{s_{i}}+\hbar\sum_{j=1}^{N}\omega_{c,e_{j}} \hat{b}_{c_{j}}^{\dagger}\hat{b}_{c_{j}}+\hbar \sum_{j=1}^{N}\omega_{r_{j},e_{j}}\hat{b}_{r_{j}}^{\dagger}\hat{b}_{r_{j}}\nonumber\\
+\hbar\sum_{j=1}^{N}(g_{s_{i}}\hat{a}_{s}^{\dagger}\hat{b}_{s_{i}}+g_{c_{j}}\hat{a}_{c}^{\dagger}\hat{b}_{c_{j}}+g_{r_{j}}\hat{a}_{r_{j}}^{\dagger}\hat{b}_{r_{j}}+h.c.)+\hbar J\sum_{j=1}^{N}\hat{a}_{c}^{\dagger}(\hat{a}_{s}+\hat{a}_{r_{j}}+h.c.),
\end{eqnarray}
where the index $i=1,..., N$ denotes that the field mode of the sender cavity interacts selectively only with the $i$th QD in that cavity. Also, $a_{s}^{\dagger}$ ($a_{s}$), $a_{c}^{\dagger}$ ($a_{c}$) and $a_{r_{j}}^{\dagger}$ ($a_{r_{j}}$) are the creation (annihilation) operators for the field modes of the sender, channel and the j$th$ receiver cavities with frequencies $w_{s,f}$, $w_{c,f}$ and $w_{r_{j},f_{j}}$ respectively. In the same way, $b_{s_{i}}^{\dagger}$ ($b_{s_{i}}$) is the creation
(annihilation) operator for the excitonic mode of the $i$th QD in the sender cavity with frequency $w_{s,e_{i}}$, $b_{c_{j}}^{\dagger}$ ($b_{c_{j}}$) and $b_{r_{j}}^{\dagger}$ ($b_{r_{j}}$) are the creation
(annihilation) operators for the excitonic modes of the j$th$ QD in the channel cavity and the QD in the j$th$ receiver cavity with frequencies $w_{c,e_{j}}$ and $w_{r_{j},e_{j}}$ respectively. On the other hand, for each selected ternary set denoted by the index $i$ we have always $w_{s,e_{i}}=w_{c,e_{i}}=w_{r_{i},e_{i}}$ and $g_{s_{i}}=g_{c_{i}}=g_{r_{i}}$ ($i=1,...,N$) as shown in Fig. 2. Corresponding to the cavities arrangement, the uniform coupling strength between the cavities denoted by $J$, strongly depends on both of the
geometry of the cavities and the actual overlap between adjacent fields of them. We cansider that $\hbar=1$, $\delta_{s_{i}}=\omega_{s,e_{i}}-\omega_{s,f}$, $\delta_{c_{j}}=\omega_{c,e_{j}}-\omega_{c,f}$ and $\delta_{r_{j}}=\omega_{r_{j},e_{j}}-\omega_{r_{j},f_{j}}$; where $\delta_{s_{i}}$, $\delta_{c_{j}}$ and $\delta_{r_{j}}$ are the detuning between the excitonic and the field modes in the sender, channel and each of the receiver parties respectively (Fig. 2). Now, the Heisenberg equations of motion for the annihilation operators of cavity fields and excitons can be obtained as follows
\begin{eqnarray}
&&\hspace{-50mm}\hat{\dot{a}}_{s}=-i( \omega_{s,f} \hat{a}_{s}+g_{s_{i}} \hat{b}_{s_{i}}+J\hat{a}_{c}), \nonumber\\
&&\hspace{-50mm}\hat{\dot{b}}_{s_{i}}=-i( \omega_{s,e_{i}} \hat{b}_{s_{i}}+g_{s_{i}} \hat{a}_{s}), \nonumber\\
&&\hspace{-50mm}\hat{\dot{a}}_{c}=-i\left( \omega_{c,f} \hat{a}_{c}+ \sum_{j=1}^{N}g_{c_{j}}\hat{b}_{c_{j}}+J(\hat{a}_{s}+\sum_{j=1}^{N}\hat{a}_{r_{j}})\right), \\
&&\hspace{-50mm}\hat{\dot{b}}_{c_{j}}=-i( \omega_{c,e_{j}} \hat{b}_{c_{j}}+g_{c_{j}} \hat{a}_{c}), \quad j=1, 2, 3, ..., N, \nonumber\\
&&\hspace{-50mm}\hat{\dot{a}}_{r_{j}}=-i( \omega_{r_{j},f_{j}} \hat{a}_{r_{j}}+g_{r_{j}} \hat{b}_{r_{j}}+J\hat{a}_{c}), \quad j=1, 2, 3, ..., N, \nonumber\\
&&\hspace{-50mm}\hat{\dot{b}}_{r_{j}}=-i( \omega_{r_{j},e_{j}} \hat{b}_{r_{j}}+g_{r_{j}} \hat{a}_{r_{j}}), \quad j=1, 2, 3, ..., N\nonumber.
\end{eqnarray}
By using the Ref. \cite{edwa}, this system of linear first order differential equations can be solved numerically. The solution of the (2) gives the explicit form for each of the operators, for example, $\hat{b}_{s_{i}}$ is obtained as
\begin{eqnarray}
&&\hspace{-20mm}\hat{b}_{s_{i}}(t)=u_{s,f}(t)\hat{a}_{s}(0)+u_{s,e_{i}}(t)\hat{b}_{s_{i}}(0)+u_{c,f}(t)\hat{a}_{c}(0)\nonumber\\
&&\hspace{-25mm}+\sum_{j=1}^{N}\left(u_{c,e_{j}}(t)\hat{b}_{c_{j}}(0)+u_{r_{j},f_{j}}(t)\hat{a}_{r_{j}}(0)+u_{r_{j},e_{j}}(t)\hat{b}_{r_{j}}(0)\right).
\end{eqnarray}
From the unitarity of the time evolution process, it is concluded that
\begin{eqnarray}
|u_{s,f}(t)|^{2}+|u_{s,e_{i}}(t)|^{2}+|u_{c,f}(t)|^{2}+\sum_{j=1}^{N}\left(|u_{c,e_{j}}(t)|^{2}+|u_{r_{j},f_{j}}(t)|^{2}
+|u_{r_{j},e_{j}}(t)|^{2}\right)=1,
\end{eqnarray}
which returns to the fact that the relation $[\hat{b}_{s_{i}}(t), \hat{b}_{s_{i}}^{\dagger}(t)]=1$ holds for all of times. Let us consider two coherent states $|\alpha\rangle$ and $|-\alpha\rangle$ with coherent amplitude $|\alpha|$ and the
overlap of them as $\langle-\alpha|\alpha\rangle=e^{-2|\alpha|^{2}}$. We identify the two coherent states as basis states for logical qubit as
\begin{eqnarray}
|0\rangle_{L}:=|\alpha\rangle,\qquad |1\rangle_{L}:=|-\alpha\rangle.
\end{eqnarray}
A properly normalized qubit constructed by superposition of coherent states $|\alpha\rangle$ and $|-\alpha\rangle$ called in abbreviate form as CSCQ, is given by
\begin{eqnarray}
|Q(\alpha)\rangle=\frac{1}{\sqrt{N_{\alpha}}}(\mu|\alpha\rangle+\nu|-\alpha\rangle),
\end{eqnarray}
where $N_{\alpha}=|\mu|^{2}+|\nu|^{2}+e^{-2|\alpha|^{2}}(\mu\nu^{\ast}+\mu^{\ast}\nu)$ is the normalization factor, and $\mu$ and $\nu$ are complex numbers. We expect that the scheme could provide the transfer a of typical CSCQ prepared on the excitonic mode of the QD which has interaction with the field of the sender cavity, to the desired receiver reliably.

\section{Selective transfer of a typical CSCQ}
As mentioned in the previous section, selective transferring a CSCQ from the sender of a network to an arbitrary receiver is our demanded protocol in this paper (see Fig. 1). The performance of the method depends on the restricting the dynamics of the system arbitrarily to each of the N ternary sets of QDs corresponding to finding the transmitted state in the respective receiver. For this end, we assume arbitrarily that (see Fig. 1) the $i$th QD in the sender cavity interacts with the field mode of the cavity and so we prepare a CSCQ, as $|Q(\alpha)\rangle$ in Eq. 6, on the excitonic mode of that QD as
\begin{eqnarray}
\begin{array}{c}
  \left|\psi(0)\right\rangle=\frac{1}{\sqrt{N_{\alpha}}}\left|0\right\rangle_{s,f}\otimes(\mu|\alpha\rangle+\nu|-\alpha\rangle)_{s,e_{i}}\otimes\left|0\right\rangle_{c,f}\\\\
  \otimes\left|0\right\rangle_{c,e_{1}}\otimes...\otimes\left|0\right\rangle_{c,e_{N}}\otimes\left|0\right\rangle_{r_{1},f_{1}}\otimes...\otimes\left|0\right\rangle_{r_{N},f_{N}}\otimes\left|0\right\rangle_{r_{1},e_{1}}\otimes...\otimes\left|0\right\rangle_{r_{N},e_{N}},
 \\\\
  \end{array}
\end{eqnarray}
where $|\alpha\rangle_{s,e_{i}}$ is a standard coherent state defined as
\begin{eqnarray}
|\alpha\rangle_{s,e_{i}}:=e^{\alpha \hat{b}^{\dagger}_{s_{i}}(0)-\alpha^{\ast} \hat{b}_{s_{i}}(0)}|0\rangle_{s,e_{i}}.
\end{eqnarray}
The time evolution of the initial state of the system is obtained as
\begin{eqnarray}
&&\hspace{-10mm}\left|\psi(t)\right\rangle=\hat{U}_{i}(t)\left|\psi(0)\right\rangle=\frac{1}{\sqrt{N_{\alpha}}}(\mu e^{\alpha\hat{b}^{\dagger}_{s_{i}}(t)-\alpha^{\ast} \hat{b}_{s_{i}}(t)}+\nu e^{-\alpha \hat{b}^{\dagger}_{s_{i}}(t)+\alpha^{\ast} \hat{b}_{s_{i}}(t)})\left|0\right\rangle_{s,f}\otimes\left|0\right\rangle_{s,e_{i}}\otimes\left|0\right\rangle_{c,f}\nonumber\\
&&\hspace{-0mm}\otimes\left|0\right\rangle_{c,e_{1}}\otimes...\otimes\left|0\right\rangle_{c,e_{N}}\otimes\left|0\right\rangle_{r_{1},f_{1}}\otimes...\otimes\left|0\right\rangle_{r_{N},f_{N}}\otimes\left|0\right\rangle_{r_{1},e_{1}}\otimes...\otimes\left|0\right\rangle_{r_{N},e_{N}},
\end{eqnarray}
where $\hat{U}_{i}(t)$ is the time evolution operator generated by the Hamiltonian $\hat{H}_{i}$. Note that in obtaining the Eq. 9, we have used the relations: $\hat{U}_{i}(t)\hat{O}(0)\hat{U}_{i}^{\dagger}(t)=\hat{O}(t)$ and $\hat{U}_{i}(t)\left|0\right\rangle=|0\rangle$ with
\begin{eqnarray}
\begin{array}{c}
  |0\rangle:=\left|0\right\rangle_{s,f}\otimes\left|0\right\rangle_{s,e_{i}}\otimes\left|0\right\rangle_{c,f}
 \\\\
  \otimes\left|0\right\rangle_{c,e_{1}}\otimes...\otimes\left|0\right\rangle_{c,e_{N}}\otimes\left|0\right\rangle_{r_{1},f_{1}}\otimes...\otimes\left|0\right\rangle_{r_{N},f_{N}}\otimes\left|0\right\rangle_{r_{1},e_{1}}\otimes...\otimes\left|0\right\rangle_{r_{N},e_{N}}.
\end{array}
\end{eqnarray}
As mentioned previously, the important condition for restricting the dynamics in a ternary set, leading to transfer the state to the desired party, is that the N ternary sets are completely different with each others in energy band gap and coupling strength to the field mode of cavities (see Fig. 2). So we expect that after a certain time, namely $t^{\ast}$, the prepared CSCQ at the sender party should be transferred with high fidelity to the QD at the $i$th receiver party, i. e.
\begin{eqnarray}
\begin{array}{c}
  \left|\psi(t^{\ast})\right\rangle=\frac{1}{\sqrt{N_{\alpha}}}\left|0\right\rangle_{s,f}\otimes|0\rangle_{s,e_{i}}\otimes\left|0\right\rangle_{c,f} \otimes\left|0\right\rangle_{c,e_{1}}\otimes...\otimes\left|0\right\rangle_{c,e_{N}}\\\\
  \otimes\left|0\right\rangle_{r_{1},f_{1}}\otimes...\otimes\left|0\right\rangle_{r_{N},f_{N}}
  \otimes\left|0\right\rangle_{r_{1},e_{1}}\otimes...\otimes(\mu|\alpha\rangle+\nu|-\alpha\rangle)_{r_{i},e_{i}}\otimes...\otimes\left|0\right\rangle_{r_{N},e_{N}}.
\end{array}
\end{eqnarray}
This situation explicitly depends of the fact that from Eq. 3, we should obtain, at a certain time $t^{\ast}$, the following equation as
\begin{eqnarray}
\hat{b}_{s_{i}}(t^{\ast})=\hat{b}_{r_{i}}(0),
\end{eqnarray}
which means that $|u_{r_{i},e_{i}}(t^{\ast})|^{2}=1$, i.e. the prepared state has been transferred to the $i$th receiver. By using this method, selective transfer of quantum state in the form of CSCQ is obtained numerically for systems with $N=2, 3, 4$ (see Fig. 2, 3, 4, 5, 6, 7, 8, 9, 10, 11). It is evident that during the time $t\in[0,t^{\ast}]$, we have $|u_{s,e_{i}}(t)|^{2}+|u_{r_{i},e_{i}}(t)|^{2}\simeq1$, means that the effective dynamics of the system relies within the $i$th ternary set of QDs and so the excitonic modes of other QDs are almost at their respective vacuum states. It is expected that the method can work for systems with higher $N$. On the other hand, let us consider the total population of the field mode of cavities in the network as below
\begin{eqnarray}
F(t)=|u_{s,f}(t)|^{2}+|u_{c,f}(t)|^{2}+\sum_{i=1}^{N}|u_{r_{i},f_{i}}(t)|^{2},
\end{eqnarray}
where this quantity is related directly to the average number of photons as follows
\begin{eqnarray}
\bar{n}=\sum_{i=1}^{N}\langle\psi(t)|(\hat{a}^{\dagger}_{s}\hat{a}_{s}+\hat{a}^{\dagger}_{c}\hat{a}_{c}+\hat{a}^{\dagger}_{r_{i}}\hat{a}_{r_{i}})|\psi(t)\rangle=\frac{|\alpha|^{2}\left(|\mu|^{2}+|\nu|^{2}-e^{-2|\alpha|^{2}}(\mu\nu^{\ast}+\mu^{\ast}\nu)\right)F(t)}{|\mu|^{2}+|\nu|^{2}+e^{-2|\alpha|^{2}}(\mu\nu^{\ast}+\mu^{\ast}\nu)}.
\end{eqnarray}
As it is evident from the Fig. 3, ..., 11, $F(t)$ takes a small values during the transfer process which in turn, leads to the vanishing amount of the average number of photons protecting against dissipations via the decay in the cavities \cite{iris, zhen, maje}. Consequently, we obtain an optimal dynamics for our proposed systems that not only performs the selective state transfer process reliably but also is robust due to the various dissipations.

\section{Conclusions}
We have presented a protocol for selective transfer of quantum information in the form of a general CSCQ reliably throughout a system of $N+2$ cavities. The sender and N receiver cavities are connected to the channel cavity through the uniform coupling arises from photon hopping. Three identical QDs of each of N different ternary set have been distributed in the sender, channel and one of the receiver cavities in such a way that there exist only one QD in each of the receiver cavities. The selective dynamics, without dispersion, leads to reliable transferring a CSBQ from the sender to the desired receiver. Also, since the effective dynamics of the system is always confined within only one of the N ternary sets of QDs, the field modes of cavities and the remainder excitonic modes are almost at their respective vacuum states, suppressing greatly efficient dissipation of the system during the state transfer processes.

\newpage
\textbf{Figure Captions}
\itemize{}
\item Fig. 1. Illustrating the selective transfer of a CSCQ in a cavity-QD system with two receivers. Two different ternary sets of QDs have been trapped in the sender, channel and receiver cavities. (a) Time evolution of the system under the Hamiltonian $\hat{H}_{1}$ transfers the prepared CSCQ to the $r_{1}$. (b) Time evolution under the Hamiltonian $\hat{H}_{2}$ transfers the prepared CSCQ to the $r_{2}$. If two ternary set are identical the dynamics under $\hat{H}_{1}$ and $\hat{H}_{2}$ are identical and the protocol does not work and the excitonic modes of QDs at the receives become to be entangled with each other.
\begin{figure}
\centering
\includegraphics[width=445 pt]{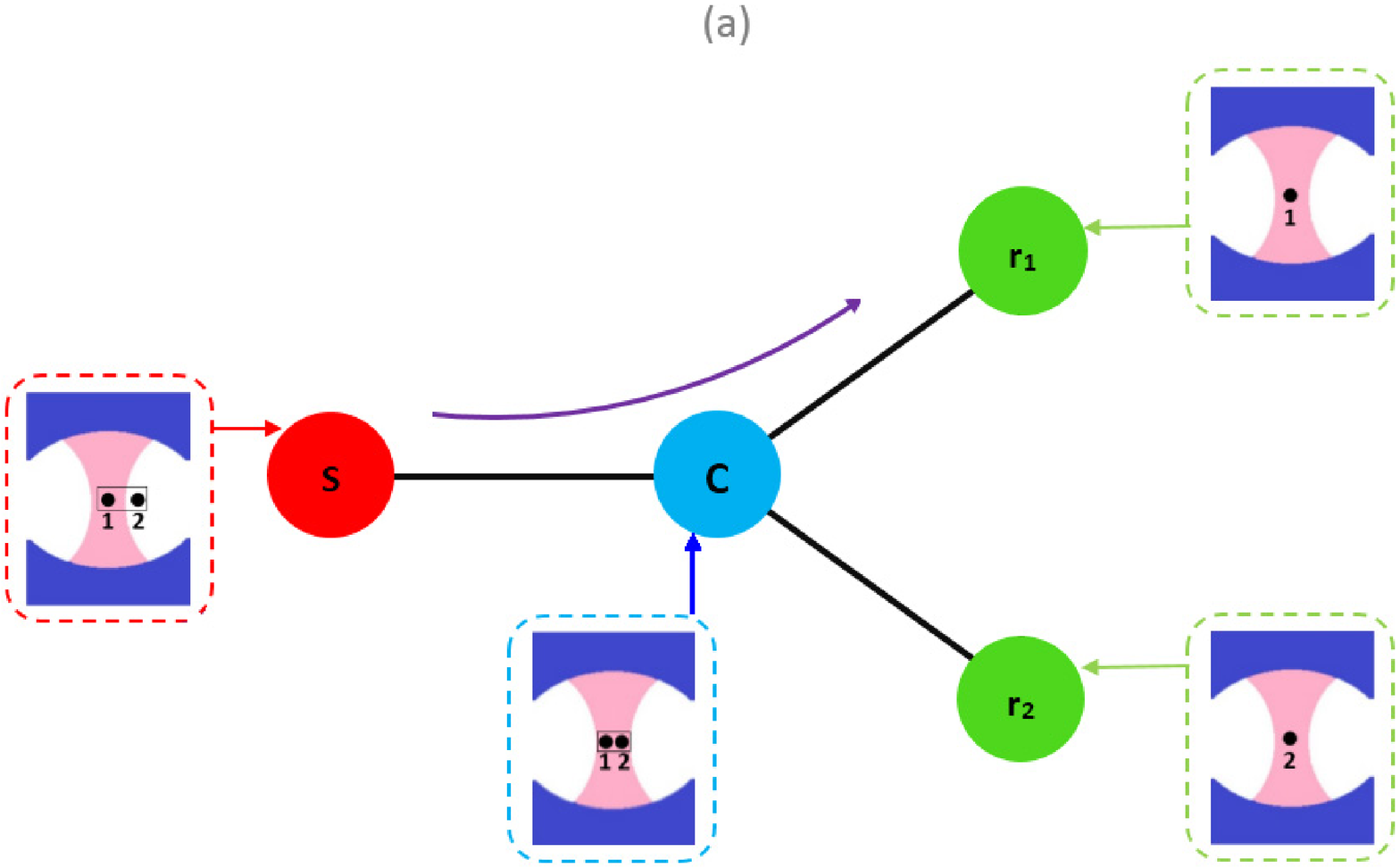}
\centering
\includegraphics[width=445 pt]{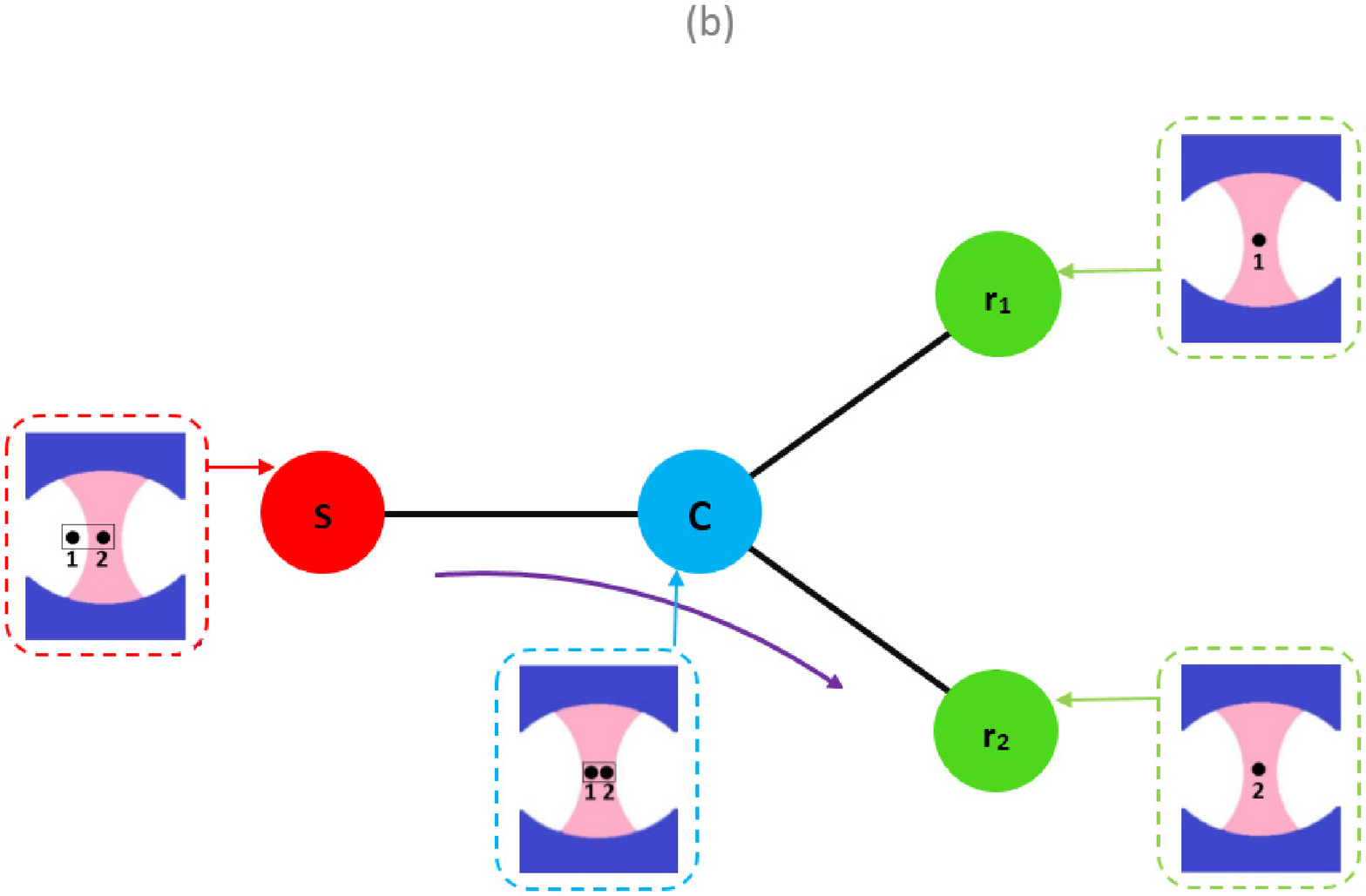}
\caption{} \label{Fig3}
\end{figure}
\newpage
\textbf{Figure Captions}
\itemize{}
\item Fig. 2. Illustration of the structure of the ternary sets of QDs. (a) The first ternary set colored by red for which $g_{s_{1}}=g_{c_{1}}=g_{r_{1}}$ and $\delta_{s_{1}}=\delta_{c_{1}}=\delta_{r_{1}}$. (b) The second ternary set colored by green for which $g_{s_{2}}=g_{c_{2}}=g_{r_{2}}$ and $\delta_{s_{2}}=\delta_{c_{2}}=\delta_{r_{2}}$. It should be noted that the relations $g_{s_{1}}\neq g_{s_{2}}$ and $\delta_{s_{1}}\neq\delta_{s_{2}}$ must be hold in such way that time evolution under $\hat{H}_{1}$ transfer the prepared CSCQ to the receiver $r_{1}$ and under $\hat{H}_{2}$ transfer the state to the receiver $r_{2}$.
\begin{figure}
\centering
\includegraphics[width=445 pt]{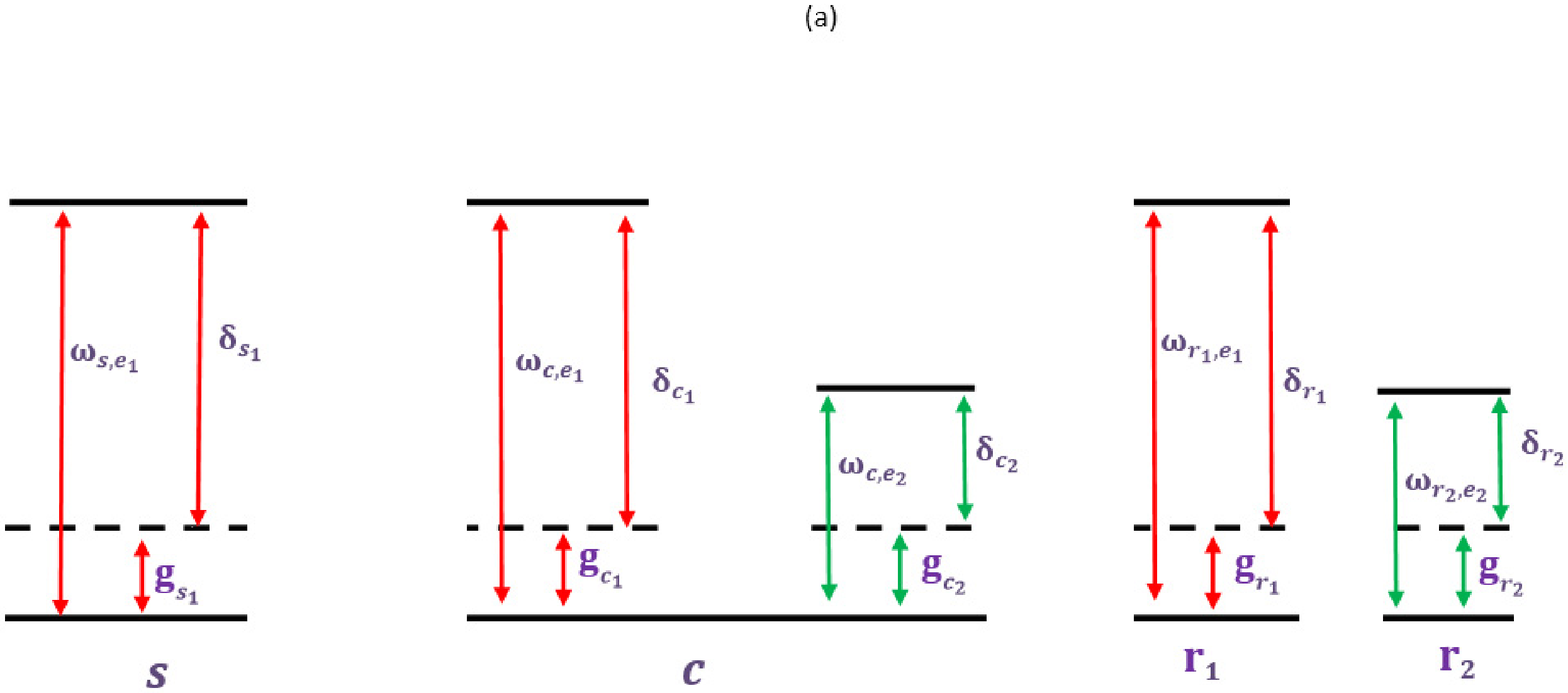}
\centering
\includegraphics[width=445 pt]{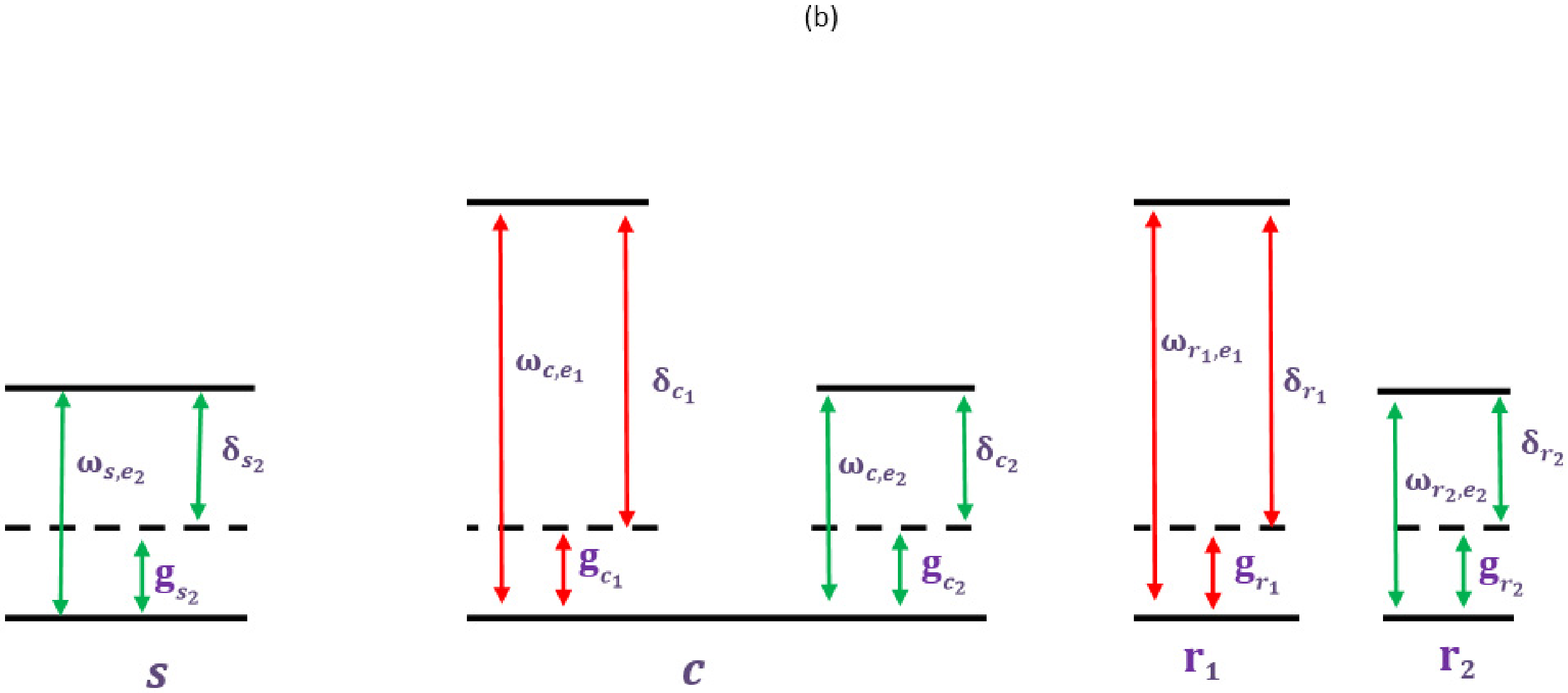}
\caption{} \label{Fig3}
\end{figure}
\newpage \newpage
\textbf{Figure Captions}
\item Fig. 3. Transferring a CSCQ from sender to the receiver $r_{1}$ for the system with $N=2$ receivers depicted in Fig. 1. The corresponding ternary set is denoted by $g_{s_{1}}=g_{c_{1}}=g_{r_{1}}=60$, $\delta_{s_{1}}=\delta_{c_{1}}=\delta_{r_{1}}=500$ (in units of $J$). $F$ is the total populations of the field modes as introduced in Eq. 13, $Us=|u_{s,e_{1}}(t)|^{2}$, $Ur_{1}=|u_{r_{1},e_{1}}(t)|^{2}$ and  $Ur_{2}=|u_{r_{2},e_{2}}(t)|^{2}$ are the populations of the excitonic modes of QDs at the sender, receivers $r_{1}$ and $r_{2}$ respectively. It is observed that at a certain time $t^{\ast}$, $Ur_{1}=|u_{r_{1},e_{1}}(t^{\ast})|^{2}\simeq1$ which means that the time evolution under the Hamiltonian $\hat{H}_{1}$ transfers the CSCQ from the sender to the receiver $r_{1}$ reliably.
    \begin{figure}
\centering
\includegraphics[width=445 pt]{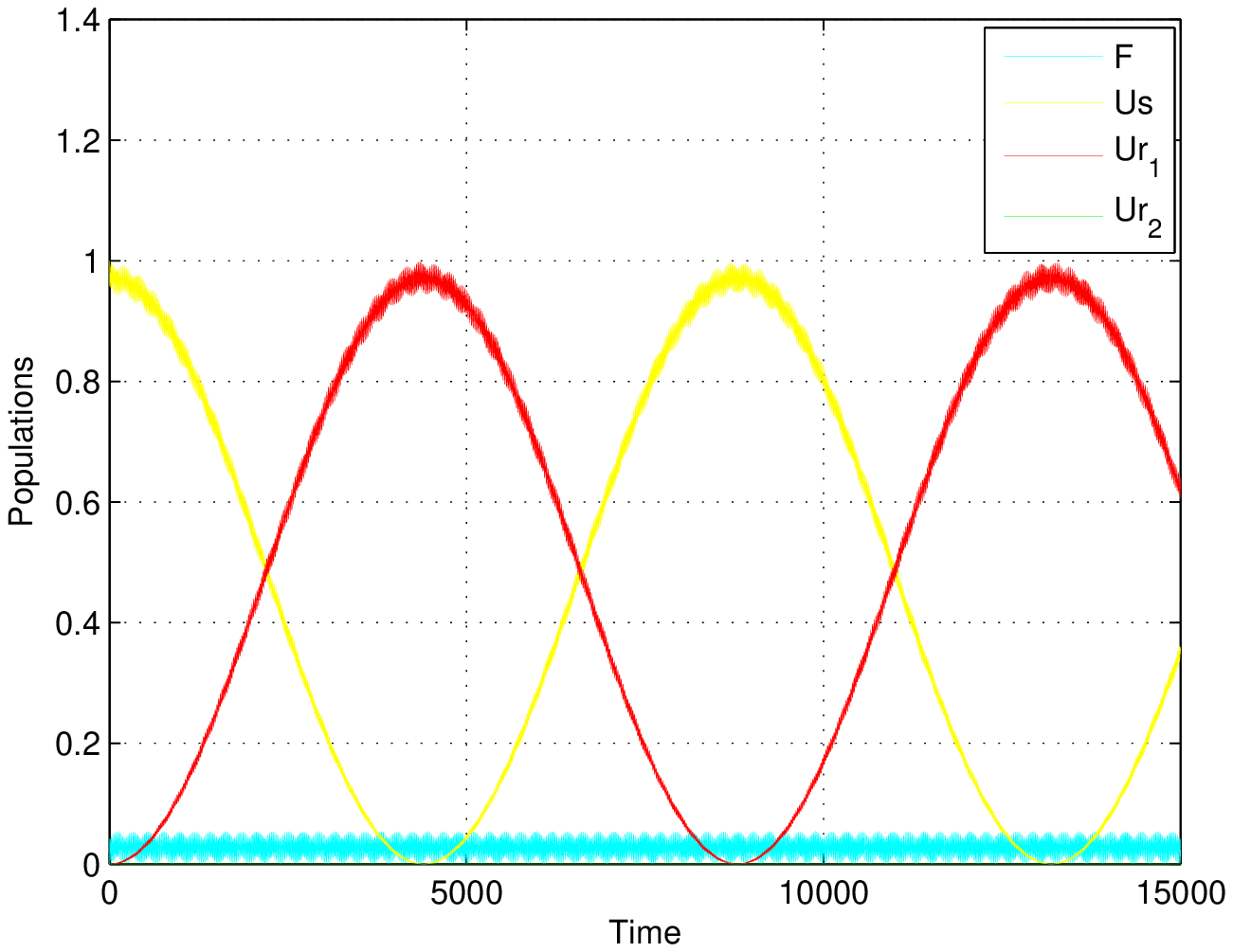}
\caption{} \label{fig1}
\end{figure}
\newpage
\textbf{Figure Captions}
\item Fig. 4. Transferring a CSCQ from sender to the receiver $r_{2}$ for the system shown in Fig. 1. The corresponding ternary set is characterized by $g_{s_{2}}=g_{c_{2}}=g_{r_{2}}=61$, $\delta_{s_{2}}=\delta_{c_{2}}=\delta_{r_{2}}=600$ (in units of $J$). At a certain time $t^{\ast}$, $Ur_{2}=|u_{r_{2},e_{2}}(t^{\ast})|^{2}\simeq1$ which means that the time evolution under the Hamiltonian $\hat{H}_{2}$ transfers the CSCQ from the sender to the receiver $r_{2}$ reliably.
     \begin{figure}
\centering
\includegraphics[width=445 pt]{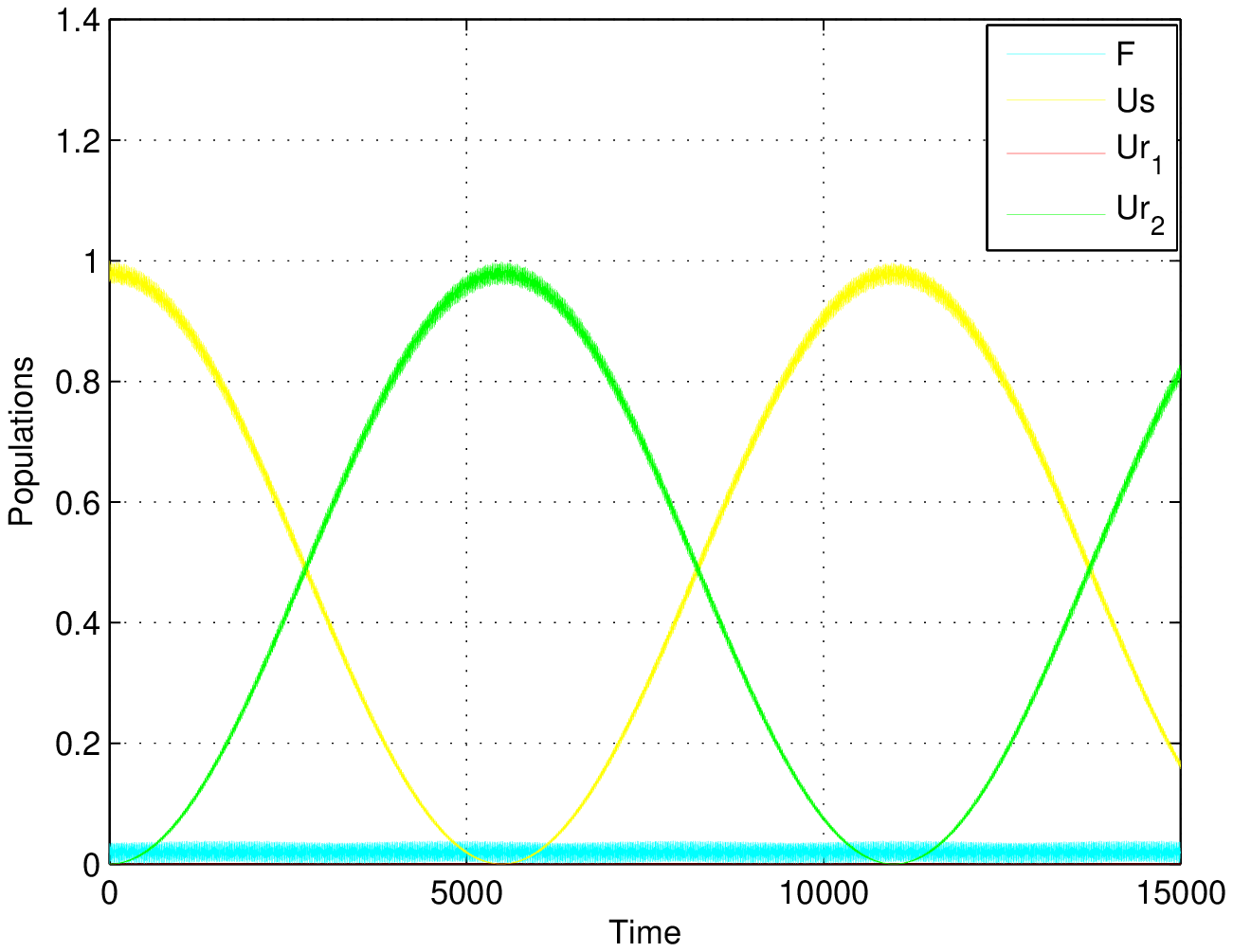}
\caption{} \label{fig2}
\end{figure}
\newpage
\textbf{Figure Captions}
\item Fig. 5. Transferring a CSCQ from sender to the receiver $r_{1}$ for a system with $N=3$ receivers. The corresponding ternary set is denoted by $g_{s_{1}}=g_{c_{1}}=g_{r_{1}}=60$, $\delta_{s_{1}}=\delta_{c_{1}}=\delta_{r_{1}}=500$ (in units of $J$). $F$ is the total populations of the field modes, $Us=|u_{s,e_{1}}(t)|^{2}$, $Ur_{1}=|u_{r_{1},e_{1}}(t)|^{2}$, $Ur_{2}=|u_{r_{2},e_{2}}(t)|^{2}$ and $Ur_{3}=|u_{r_{3},e_{3}}(t)|^{2}$ are the populations of the excitonic modes of QDs at the sender, receivers $r_{1}$, $r_{2}$ and $r_{3}$ respectively. For a certain time $t^{\ast}$, $Ur_{1}=|u_{r_{1},e_{1}}(t^{\ast})|^{2}\simeq1$ which shows that the time evolution under the Hamiltonian $\hat{H}_{1}$ transfers the CSCQ from the sender to the receiver $r_{1}$ reliably.
\begin{figure}
\centering
\includegraphics[width=445 pt]{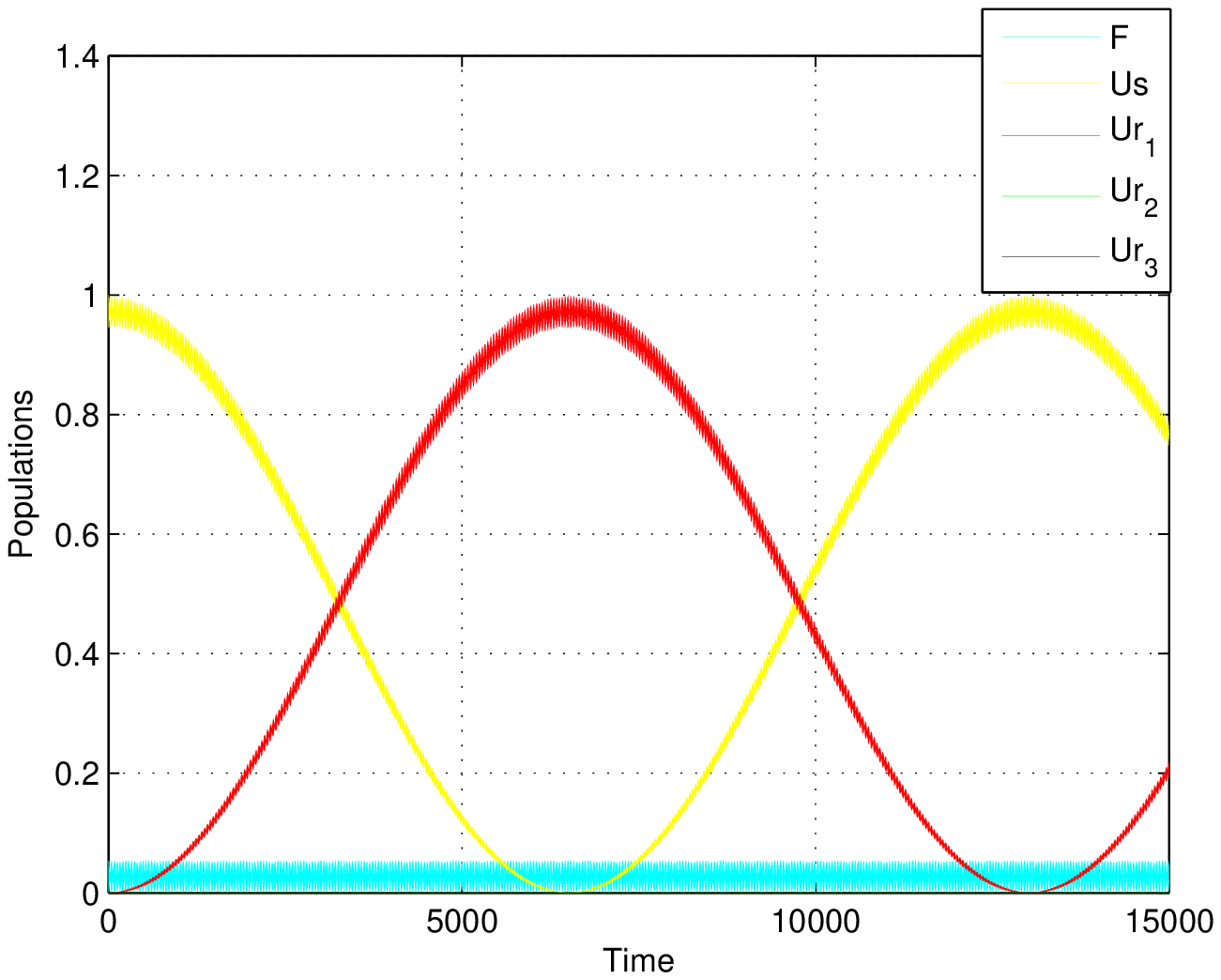}
\caption{} \label{fig}
\end{figure}
\newpage
\textbf{Figure Captions}
\item Fig. 6. Transferring a CSCQ from sender to the receiver $r_{2}$ for the system with $N=3$ receivers. The corresponding ternary set is denoted by $g_{s_{2}}=g_{c_{2}}=g_{r_{2}}=61$, $\delta_{s_{2}}=\delta_{c_{2}}=\delta_{r_{2}}=600$ (in units of $J$). For a certain time $t^{\ast}$, $Ur_{2}=|u_{r_{2},e_{2}}(t^{\ast})|^{2}\simeq1$ which shows that the time evolution under the Hamiltonian $\hat{H}_{2}$ transfers the CSCQ from the sender to the receiver $r_{2}$ reliably.
\begin{figure}
\centering
\includegraphics[width=445 pt]{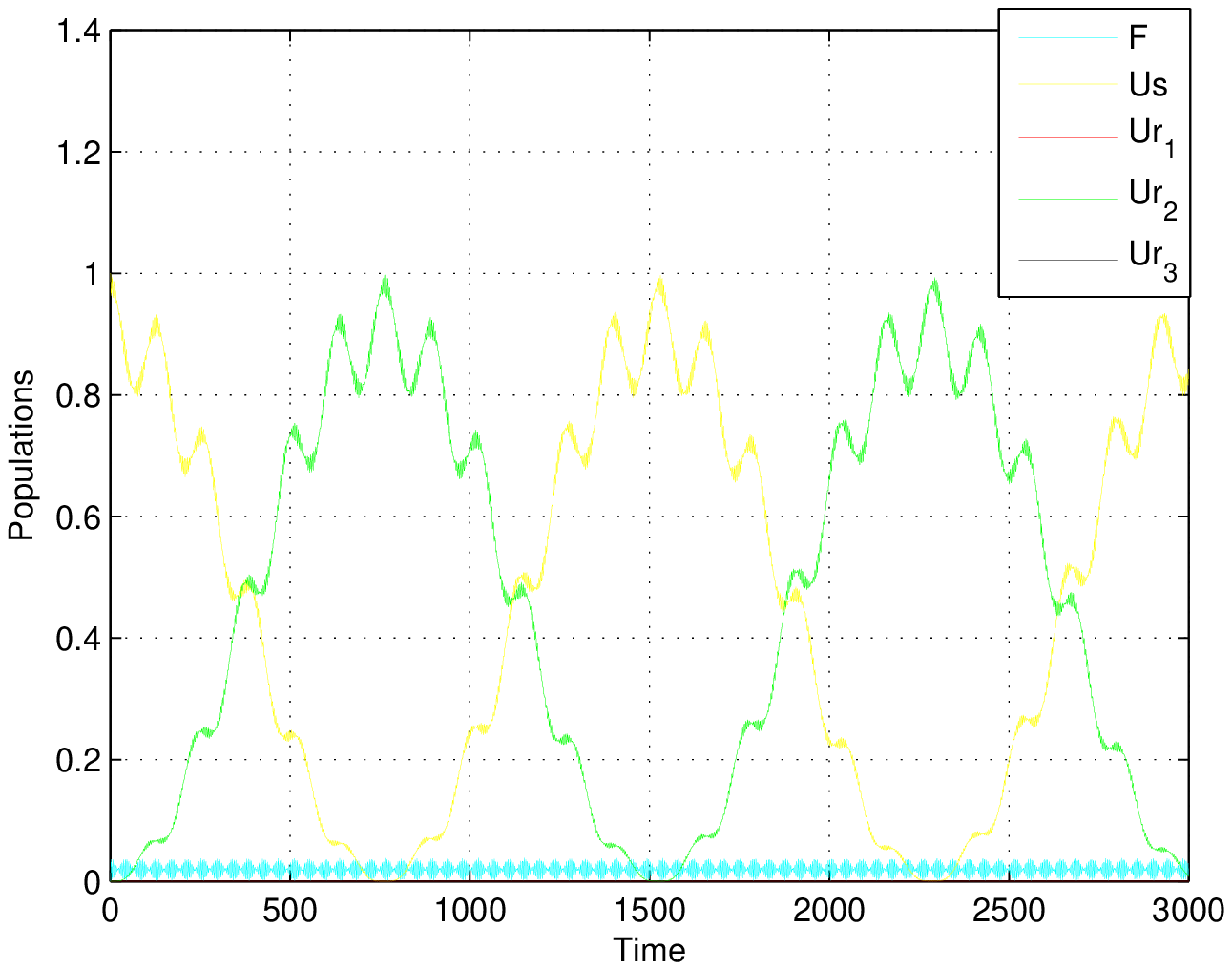}
\caption{} \label{fig2}
\end{figure}
\newpage
\textbf{Figure Captions}
\item Fig. 7. Transferring a CSCQ from sender to the receiver $r_{3}$ for the system with $N=3$ receivers. The corresponding ternary set is denoted by $g_{s_{3}}=g_{c_{3}}=g_{r_{3}}=62$, $\delta_{s_{3}}=\delta_{c_{3}}=\delta_{r_{3}}=700$ (in units of $J$). At time $t^{\ast}$, $Ur_{3}=|u_{r_{3},e_{3}}(t^{\ast})|^{2}\simeq1$ which shows that the time evolution under the Hamiltonian $\hat{H}_{3}$ transfers the CSCQ from the sender to the receiver $r_{3}$ reliably.
\begin{figure}
\centering
\includegraphics[width=445 pt]{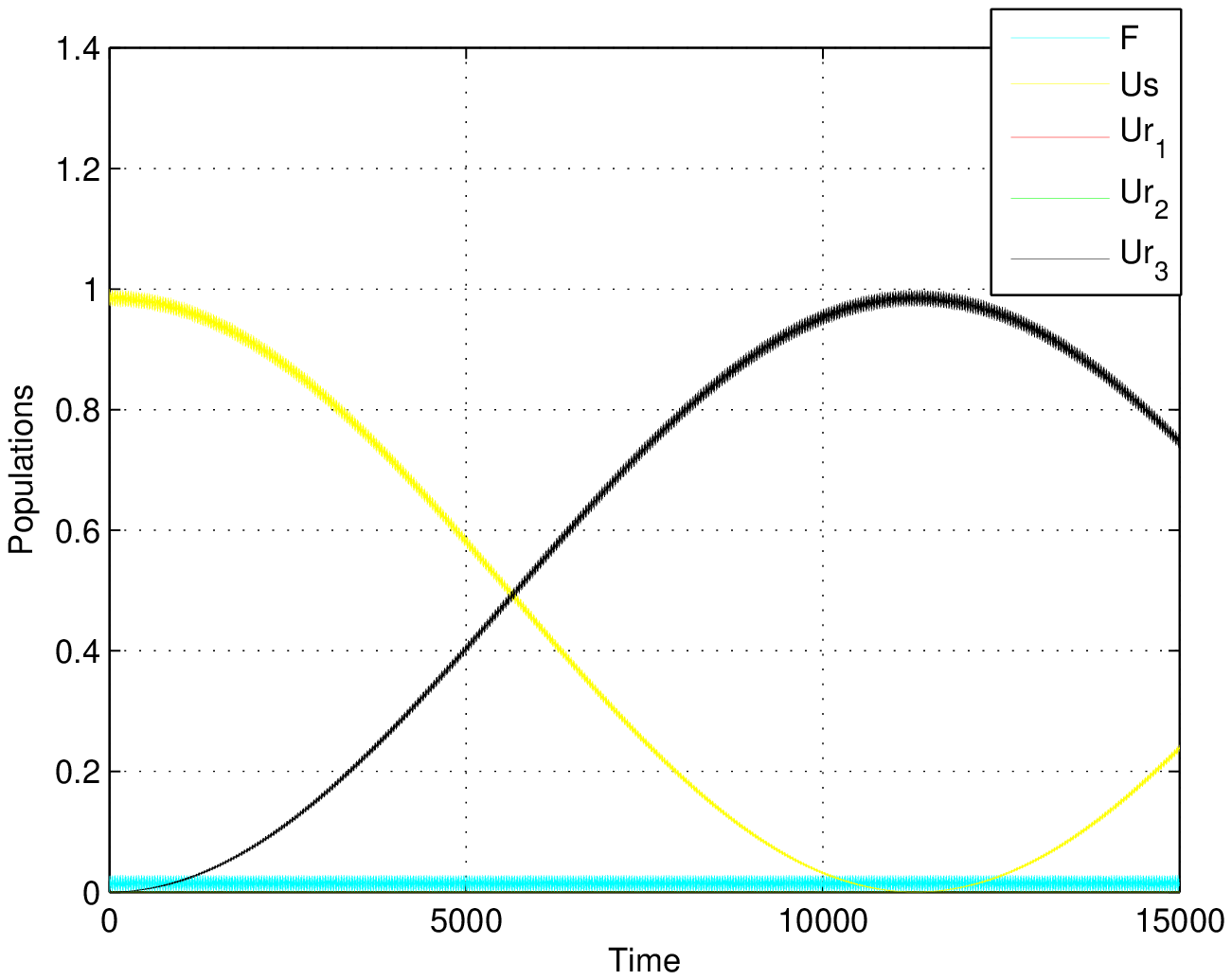}
\caption{} \label{fig2}
\end{figure}
\newpage
\textbf{Figure Captions}
\item Fig. 8. Transferring a CSCQ from sender to the receiver $r_{1}$ for a system with $N=4$ receivers. The corresponding ternary set is denoted by $g_{s_{1}}=g_{c_{1}}=g_{r_{1}}=60$, $\delta_{s_{1}}=\delta_{c_{1}}=\delta_{r_{1}}=500$ (in units of $J$). $F$ is the total populations of the field modes, $Us=|u_{s,e_{3}}(t)|^{2}$, $Ur_{1}=|u_{r_{1},e_{1}}(t)|^{2}$, $Ur_{2}=|u_{r_{2},e_{2}}(t)|^{2}$, $Ur_{3}=|u_{r_{3},e_{3}}(t)|^{2}$ and $Ur_{4}=|u_{r_{4},e_{4}}(t)|^{2}$ are the populations of the excitonic modes of QDs at the sender and receivers $r_{1}$, $r_{2}$, $r_{3}$ and $r_{4}$ respectively. At the time $t^{\ast}$, $Ur_{1}=|u_{r_{1},e_{1}}(t^{\ast})|^{2}\simeq1$ which certifies that the time evolution under the Hamiltonian $\hat{H}_{1}$ transfers the CSCQ from the sender to the receiver $r_{1}$.
\begin{figure}
\centering
\includegraphics[width=445 pt]{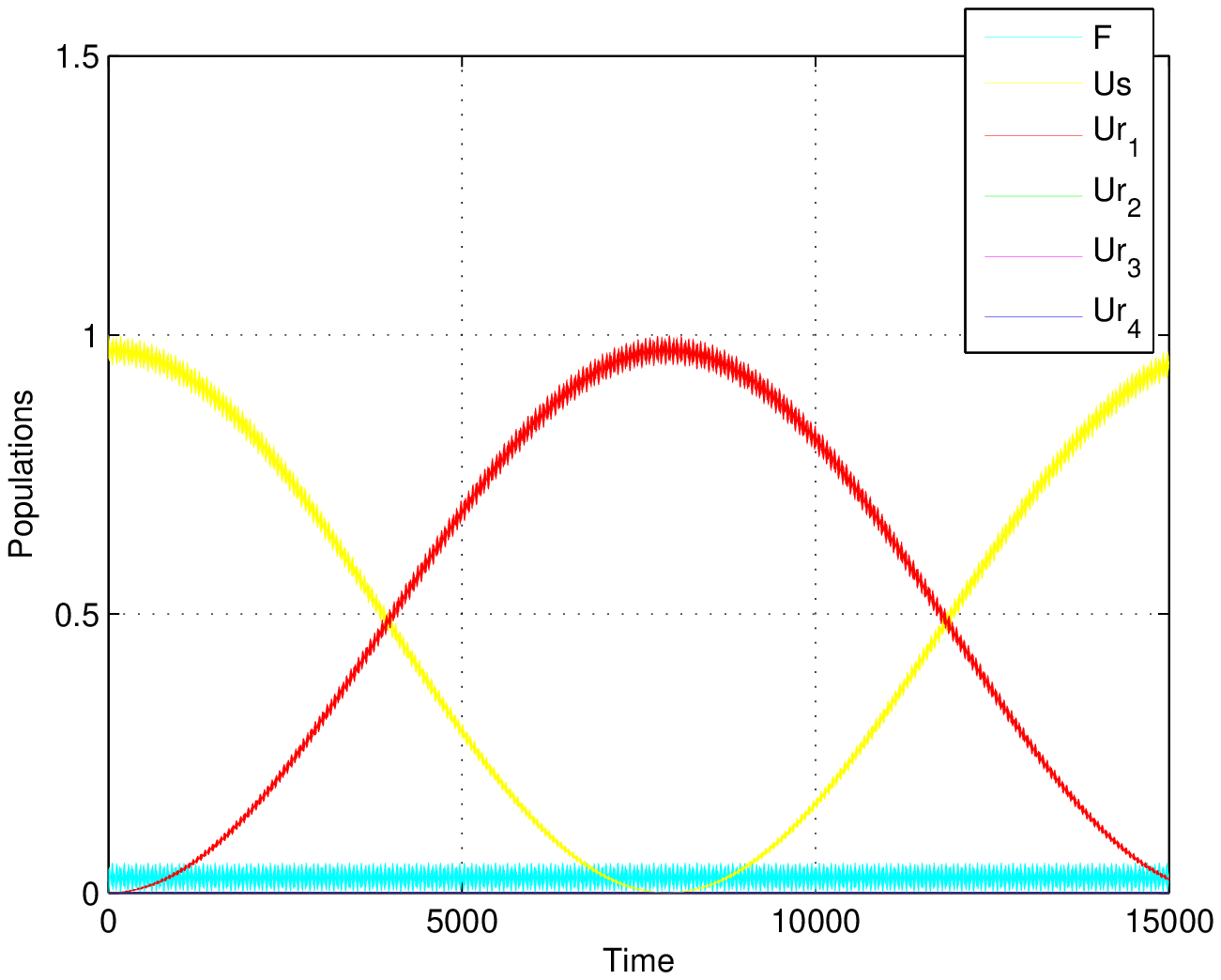}
\caption{} \label{fig2}
\end{figure}
\newpage
\textbf{Figure Captions}
\item Fig. 9. Transferring a CSCQ from sender to the receiver $r_{2}$ for the system with $N=4$ receivers. The corresponding ternary set is denoted by $g_{s_{2}}=g_{c_{2}}=g_{r_{2}}=61$, $\delta_{s_{2}}=\delta_{c_{2}}=\delta_{r_{2}}=600$ (in units of $J$). At the time $t^{\ast}$, $Ur_{2}=|u_{r_{2},e_{2}}(t^{\ast})|^{2}\simeq1$ which indicates that the time evolution under the Hamiltonian $\hat{H}_{2}$ transfers the CSCQ from the sender to the receiver $r_{2}$ reliably.
\begin{figure}
\centering
\includegraphics[width=445 pt]{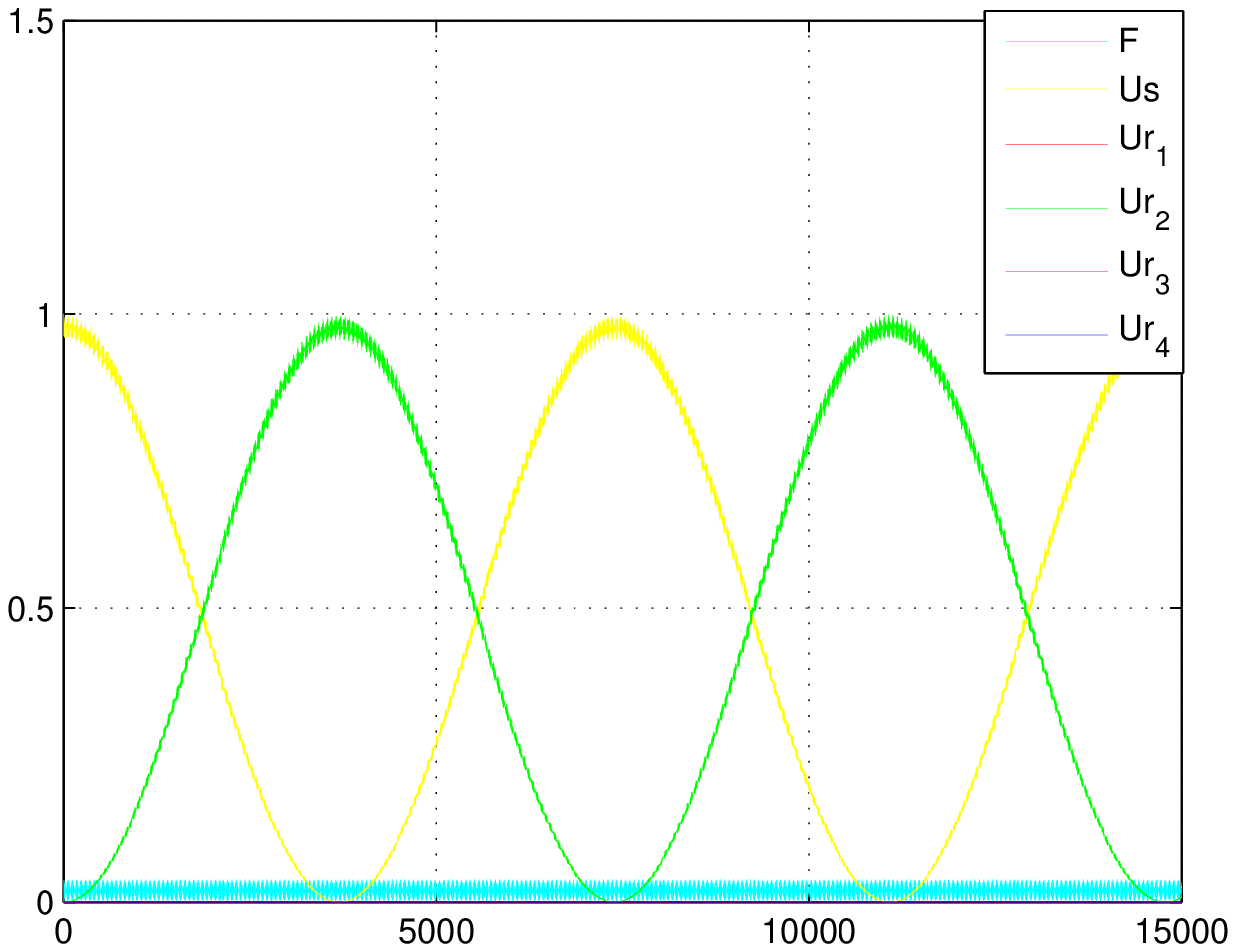}
\caption{} \label{fig2}
\end{figure}
\newpage
\textbf{Figure Captions}
\item Fig. 10. Transferring a CSCQ from sender to the receiver $r_{3}$ for the system with $N=4$ receivers. The corresponding ternary set is denoted by $g_{s_{3}}=g_{c_{3}}=g_{r_{3}}=62$, $\delta_{s_{3}}=\delta_{c_{3}}=\delta_{r_{3}}=700$ (in units of $J$). At $t^{\ast}$, $Ur_{3}=|u_{r_{3},e_{3}}(t^{\ast})|^{2}\simeq1$ which grantees that the time evolution under the Hamiltonian $\hat{H}_{3}$ transfers the CSCQ from the sender to the receiver $r_{3}$.
\begin{figure}
\centering
\includegraphics[width=445 pt]{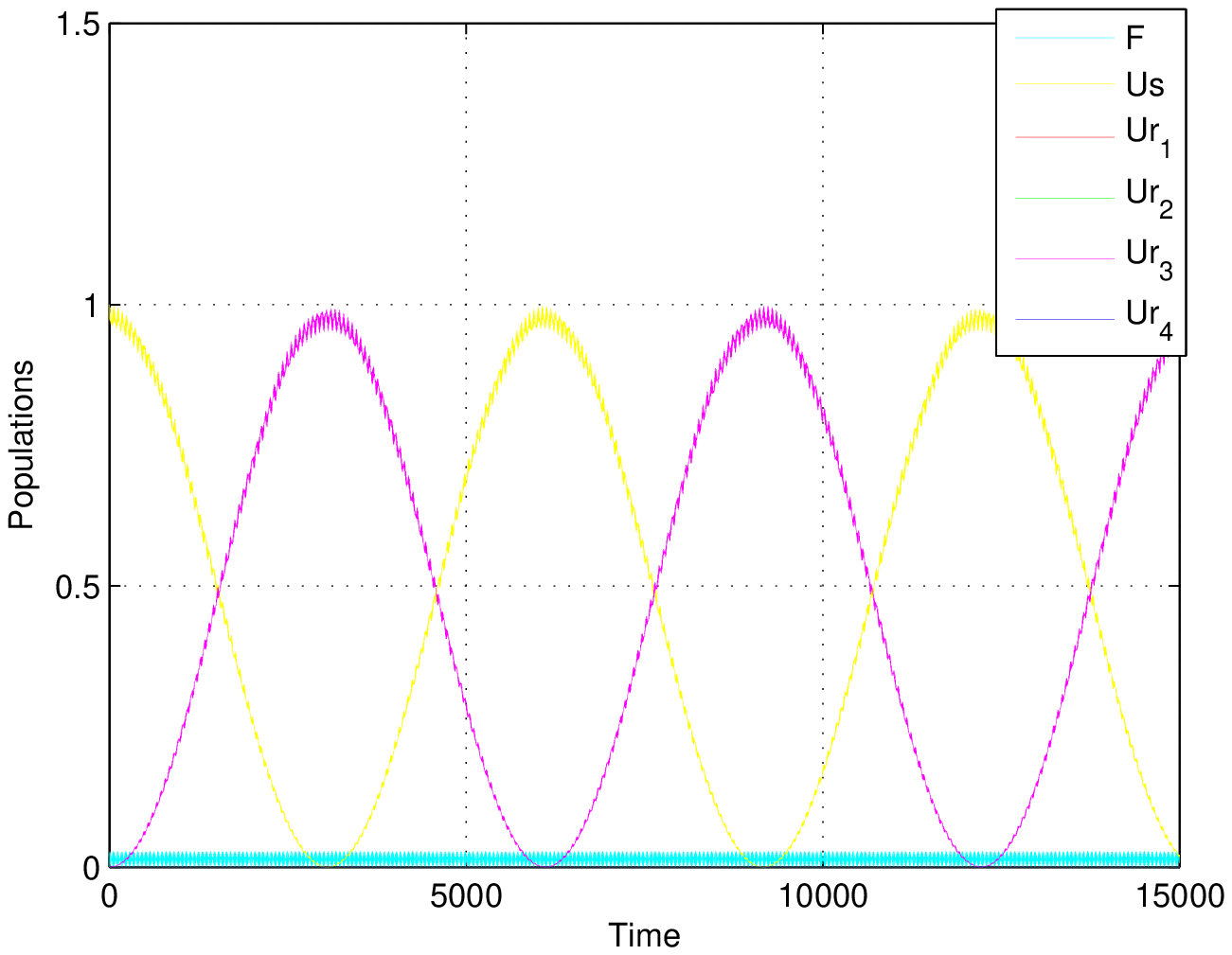}
\caption{} \label{fig2}
\end{figure}
\newpage
\textbf{Figure Captions}
\item Fig. 11. Transferring a CSCQ from sender to the receiver $r_{4}$ for the system with $N=4$ receivers. The corresponding ternary set is denoted by $g_{s_{4}}=g_{c_{4}}=g_{r_{4}}=63$, $\delta_{s_{4}}=\delta_{c_{4}}=\delta_{r_{4}}=800$ (in units of $J$). At time $t^{\ast}$, $Ur_{4}=|u_{r_{4},e_{4}}(t^{\ast})|^{2}\simeq1$ which ensures that the time evolution under the Hamiltonian $\hat{H}_{4}$ transfers the CSCQ from the sender to the receiver $r_{4}$ reliably.
\begin{figure}
\centering
\includegraphics[width=445 pt]{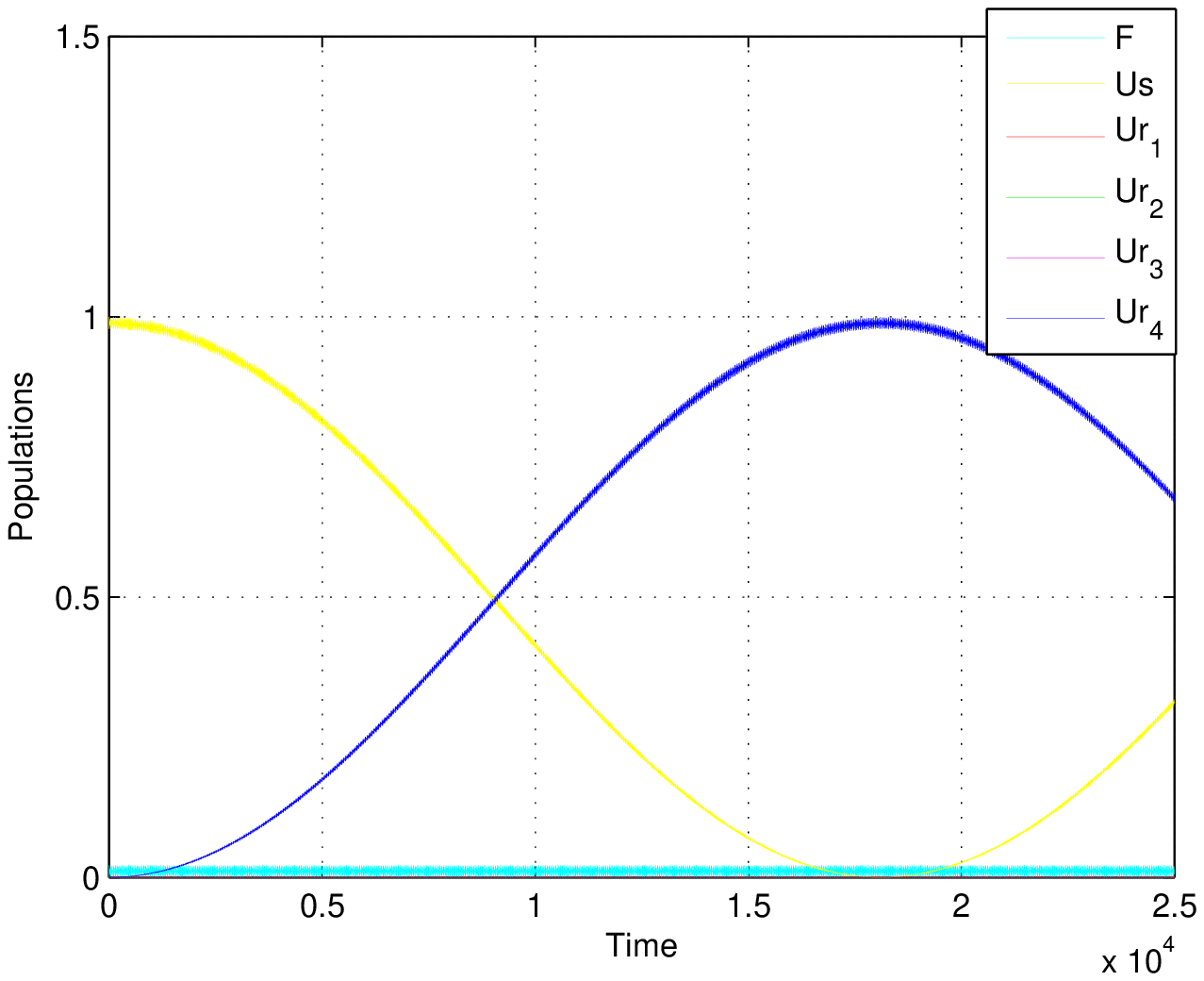}
\caption{} \label{fig2}
\end{figure}
\end{document}